\documentclass[letterpaper]{article}
\PassOptionsToPackage{numbers, compress}{natbib}
\usepackage[preprint]{neurips_2025}
\usepackage{times} 
\usepackage{helvet} 
\usepackage{array}
\usepackage{courier} 
\usepackage[hyphens]{url} 
\usepackage{graphicx} 
\usepackage{graphicx} 
\usepackage{natbib} 
\usepackage{caption} 
\usepackage{amsmath, amssymb, graphicx, booktabs} 
\usepackage{algorithm, algpseudocode}
\usepackage{subcaption} 
\usepackage{multirow} 
\usepackage{hyperref}
\title{Decoding EEG Speech Perception with Transformers and VAE-based Data Augmentation}

\author{
  Terrance Yu-Hao Chen \\
  Carnegie Mellon University \\
  Pittsburgh, PA, USA \\
  \texttt{terrancc@andrew.cmu.edu}
  \And
  Yulin Chen \\
  Carnegie Mellon University \\
  Pittsburgh, PA, USA \\
  \texttt{jolinc@andrew.cmu.edu}
  \And
  Pontus Soederhaell \\
  Carnegie Mellon University \\
  Pittsburgh, PA, USA \\
  \texttt{psoderha@andrew.cmu.edu}
  \And
  Sadrishya Agrawal \\
  Carnegie Mellon University \\
  Pittsburgh, PA, USA \\
  \texttt{sadrisha@andrew.cmu.edu}
  \And
  Kateryna Shapovalenko \\
  Carnegie Mellon University \\
  Pittsburgh, PA, USA \\
  \texttt{kshapova@alumni.cmu.edu}
}

\begin{document}

\maketitle

\begin{abstract}
Decoding natural speech from non-invasive electroencephalography (EEG) remains a fundamental challenge for practical brain–computer interfaces (BCIs) due to low signal-to-noise ratio, limited labeled data, and high inter-subject variability. In this work, we present a hybrid EEG-to-speech decoding framework that adapts a state-of-the-art electromyography (EMG) speech decoder to the EEG modality and extends it with two training objectives: a word-level classifier and a sequence-to-sequence (Seq2Seq) decoder. To mitigate data scarcity, we implement a variational autoencoder (VAE)–based EEG augmentation module that generates synthetic signals in latent space to improve model robustness. Evaluated on a publicly available EEG dataset of speech perception, our framework demonstrates the feasibility of capturing word- and sentence-level linguistic dynamics from EEG. All code and models are released open-source to establish a reproducible baseline for future EEG-to-text research. This work represents one of the first systematic adaptations of EMG-to-speech modeling to EEG and a step toward scalable, non-invasive brain-to-text systems.
\end{abstract}

\section{Introduction}
\label{sec:introduction}
Surface electroencephalography (EEG) is a widely used non-invasive method for monitoring neural activity, offering millisecond-level temporal resolution with minimal setup costs. With recent advances in deep learning, EEG-based brain–computer interfaces (BCIs) have progressed from basic motor control to more complex cognitive tasks such as speech decoding. However, EEG decoding presents several persistent challenges: (1) low signal-to-noise ratio (SNR), (2) strong inter-subject variability, and (3) limited availability of labeled, high-quality datasets for training data-intensive models. To address these challenges, we propose a two-pronged approach:
\begin{itemize}
    \item \textbf{Variational Autoencoders (VAEs)}: We use VAEs to learn latent representations of EEG and EMG signals, enabling the generation of diverse and realistic synthetic samples for training augmentation \cite{CAI2024106131,chien2022maeegmaskedautoencodereeg}.
    \item \textbf{Cross-Modal Transformer Adaptation}: We adapt a state-of-the-art EMG-based speech decoding transformer for EEG input, enabling cross-modal knowledge transfer and improving decoding performance on sentence-level tasks.
\end{itemize}

We validate our method on the public dataset, which contains continuous speech perception recordings from multiple subjects. Our findings demonstrate that generative augmentation and transformer-based modeling improve generalization and accuracy in EEG-based speech decoding. This work contributes to the development of scalable, modality-agnostic BCI systems for real-world language decoding applications.

\section{Literature Review}

Brain-to-text decoding is an emerging area in neuroprosthetics and brain–computer interfaces (BCIs), with high potential for restoring communication in individuals affected by neurological disorders. This technology aims to interpret brain activity related to speech (perceived, spoken, or imagined), enabling users to communicate directly through neural signals. The principal paradigms explored in speech decoding BCIs include speech perception, overt speech, silent speech, and inner or imagined speech.

\subsection{Speech Perception}
Speech perception involves decoding brain activity while a subject listens to external speech stimuli. Although this passive paradigm does not reflect voluntary communication, it provides foundational insights into auditory processing and the neural mechanisms underlying language comprehension. Recent work has employed artificial intelligence to decode brain activity during natural speech perception. One study explored transformer-based decoding while subjects listened to multi-speaker dialogues \cite{meta}, revealing fine-grained temporal correlations between EEG and linguistic structure. 

\subsection{Speech Production}

\subsubsection{Active/Overt Speech}
Overt speech decoding focuses on interpreting brain activity during actual vocalization. Despite the introduction of muscular artifacts from speech-related motion, this paradigm provides direct mappings between speech content and neural signals. Recent multimodal approaches have fused EEG with audio recordings to improve automatic speech recognition (ASR) under adverse conditions. One model integrated EEG and audio features using deep learning and achieved 95.39\% classification accuracy, outperforming audio-only baselines in white-noise environments \cite{DAS2024}. These results demonstrate that EEG signals carry complementary information that can enhance ASR robustness in noisy or occluded speech scenarios.

\subsubsection{Silent Speech}
Silent speech decoding targets the articulatory motor signals generated during speech preparation, such as tongue, lip, or jaw movement—without vocalization. This direction is particularly promising for individuals with vocal impairments, as it circumvents the need for audible output. High-resolution intracranial systems have achieved promising results: for instance, a recent ECoG-based decoder reconstructed full words with 94\% accuracy using phoneme-level alignment \cite{highperformance_speech}. Other systems integrate EMG and lip-reading sensors, combining complementary motor signals to improve performance across environments and users \cite{DENBY2010270,gaddy2020digital}. Cross-modal models like MONA (Multimodal Orofacial Neural Audio) incorporate both neural and acoustic data, yielding substantial reductions in word error rates across constrained-vocabulary tasks \cite{crossmodal}.

\subsubsection{Imaginary/Inner Speech}
Imagined speech refers to the mental simulation of speaking without any articulatory motion, while inner speech captures the subjective experience of “thinking in words.” Though often used interchangeably, these paradigms differ in experimental setup and mental state evocation. Several studies have investigated decoding imagined speech using EEG. In one study, subjects imagined directional commands (“up,” “down,” etc.), which were classified using long short-term memory (LSTM) networks and achieved 92.5\% accuracy across four classes using an 8-channel EEG system \cite{LSTMRNN_EEGClassification}. Another study evaluated classification performance across 5–6 classes using Random Forests and Support Vector Machines, reporting lower accuracies in the 18–22\% range, likely due to increased label granularity and limited subject adaptation \cite{German_etal_2017}. Performance differences between phonemes, short words, and long words were also explored in work using Relevance Vector Machines \cite{Nguyen_2018}. Classification accuracy increased with linguistic complexity: 49.0\% for vowels, 51.1\% for short words, and 66.2\% for long words. These results suggest that longer linguistic units may yield more discriminable EEG patterns, motivating the design of sentence-level imagined speech decoding pipelines.

\section{Data}
\subsection{Dataset}
\label{dataset}
We use the publicly available dataset, which includes EEG recordings from 33 adult volunteers (after exclusions) \cite{brennan2019hierarchical}. Participants passively listened to a 12.4-minute audiobook narration of the first chapter of \textit{Alice's Adventures in Wonderland}, slowed by 20\% to aid comprehension. The narrative consists of 2,129 words grouped into 84 sentences. EEG signals were recorded using 61 active electrodes at a 500 Hz sampling rate, with a 0.1–200 Hz bandpass filter.

\subsection{Data Preprocessing}
\label{preprocessing}

EEG signals are highly susceptible to noise from artifacts such as eye movements, muscle activity, and environmental interference, leading to a low signal-to-noise ratio (SNR) \cite{Shamlo2015The}. To address this, we designed a preprocessing pipeline that produces two types of EEG representations:

\subsubsection{Minimally processed EEG data (\texttt{eeg\_raw})}

We applied the following steps: \textbf{channel removal} (the last two EEG channels were discarded); \textbf{baseline correction} (the mean of the first 0.5 seconds was subtracted to eliminate DC drift); \textbf{robust scaling} (outliers were reduced using scikit-learn’s robust scaler); \textbf{outlier handling} (values outside the 5th–95th percentile range were clipped, and extreme values exceeding 20 standard deviations were clamped); \textbf{standardization} (the data was normalized to zero mean and unit variance). 

\subsubsection{Feature-enhanced EEG representations (\texttt{eeg\_feats})}

This version includes additional transformations: \textbf{temporal shifting} (signals were shifted by 150 ms to account for neural response delays); \textbf{feature extraction} (using convolutional operations, we computed the following: double-averaged signal, RMS of wavelet coefficients and rectified signal, zero-crossing rate, and mean of the rectified signal); \textbf{feature stacking} (all extracted features were concatenated to form the final representation).

\section{Model Description}

\subsection{Baseline EEG-to-Speech Model}
\label{sec:baseline-model}

\begin{figure}
    \centering
    \includegraphics[width=0.6\linewidth]{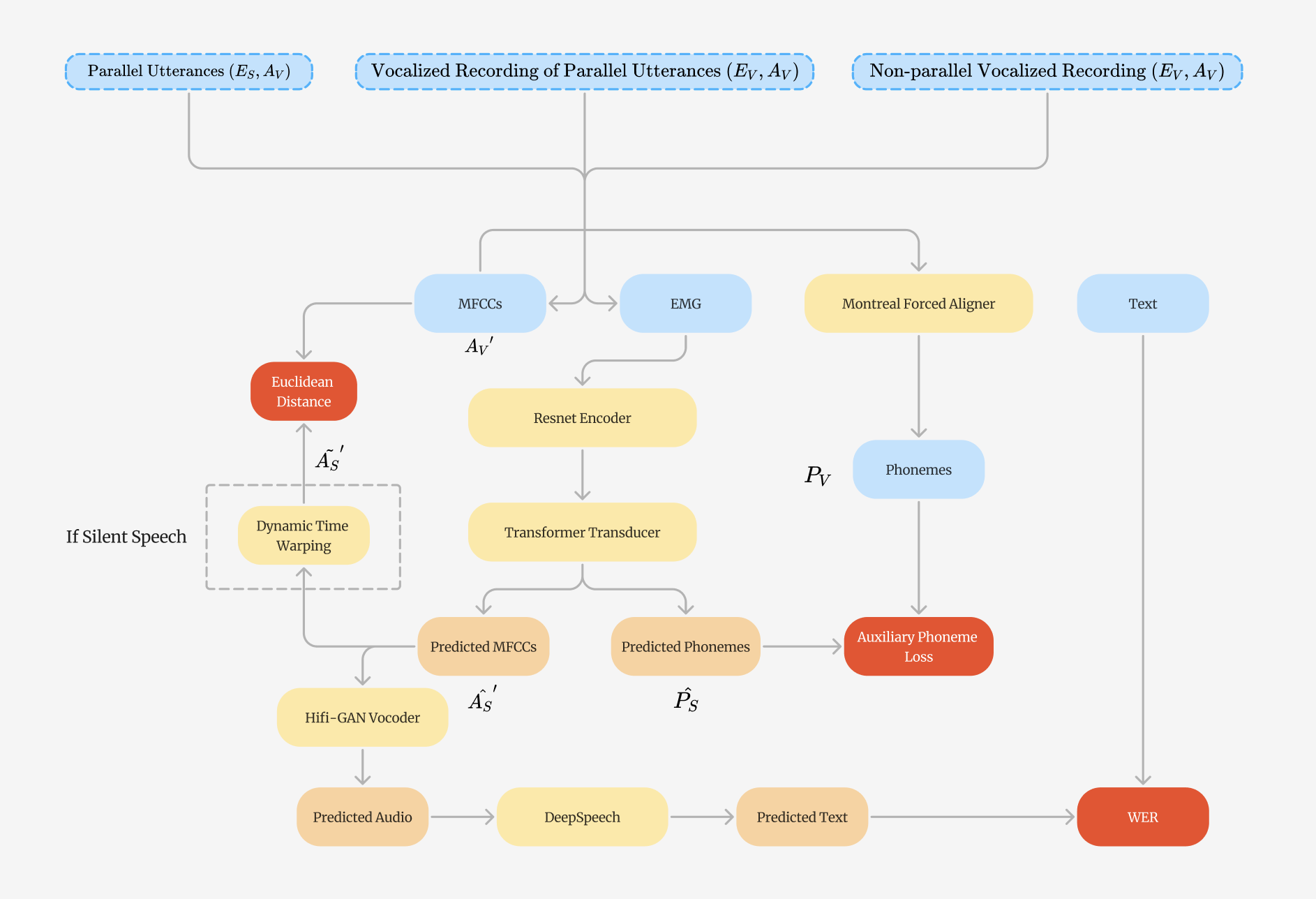}
    \caption{Baseline model.}
    \label{fig:baseline-model-overflow}
\end{figure}

We base our work on the state-of-the-art EMG-to-speech model from \textit{Voicing Silent Speech} \cite{Gaddy_Dan_2022}, one of the first deep learning system trained specifically to convert silently mouthed EMG signals into audible speech. This model also serves as the foundation for \textit{A Cross-Modal Approach to Silent Speech with LLM-Enhanced Recognition} \cite{crossmodal} and has become a benchmark for silent speech decoding. One of its key innovations is a cross-modal training pipeline that aligns EMG signals from silent speech with audio from vocalized speech, a necessary workaround given the lack of audio in silent speech data. The model combines CNN-based feature extraction, Transformer-based sequence transduction, alignment via Dynamic Time Warping (DTW), and speech synthesis via HiFi-GAN (Figure~\ref{fig:baseline-model-overflow}). We adapt this architecture to EEG-based decoding by modifying its feature extractor and retraining it on the dataset from \cite{brennan2019hierarchical}.

\vspace{0.5em}
The model consists of the following components:

\begin{itemize}
    \item \textbf{Feature Extraction:} Raw EMG (or EEG) signals are preprocessed and passed through a CNN with three residual blocks (2×Conv3 + BN + ReLU + shortcut Conv1) to produce a latent feature representation $E'_S$. These features are then passed to a transduction model to predict audio features $\hat{A}'_S$ (e.g., mel-spectrograms, MFCC, phonemes).
    
    \item \textbf{Transduction via Transformer:} The core transduction model is a Transformer that learns temporal alignments via multi-head self-attention. A learned relative position embedding $p_{ij}$ is added to the key vector:
    \[
    a_{ij} = \text{softmax} \left( \frac{(W_K x_j + p_{ij})^\top (W_Q x_i)}{\sqrt{d}} \right)
    \]
    where $W_K$, $W_Q$ are learned projections and $d$ is the dimensionality.
    
    \item \textbf{Cross-Modal Training:} For vocalized data, the model minimizes the Euclidean loss between predicted and ground truth audio features:
    \[
    \mathcal{L}_{\text{voc}} = \| \hat{A}'_V - A_V \|^2
    \]
    For silent data, Dynamic Time Warping (DTW) aligns predictions to the corresponding vocalized audio:
    \[
    \delta[i,j]=\|\hat{A}'_S[i]-A'_V[j]\|
    \]
    
    \item \textbf{Auxiliary Phoneme Loss:} A softmax layer predicts framewise phonemes. Phoneme labels $P$ are obtained via Montreal Forced Aligner. The total loss becomes:
    \[
    \mathcal{L} = \sum_i \left\| A'[i] - \tilde{A}[i] \right\|^2 + \lambda P[i]^\top \log \tilde{P}[i]
    \]
    
    \item \textbf{Vocoding:} Final mel-spectrogram predictions are converted to audio using HiFi-GAN \cite{hifigan}, a parallel neural vocoder trained on vocalized speech.
\end{itemize}

\vspace{0.5em}
To adapt the EMG-based architecture to the EEG decoding task, we made two primary changes:

\begin{enumerate}
    \item We modified the feature extraction pipeline to handle EEG-specific preprocessing and input representations (see Sections~\ref{sec:word-classifier} and~\ref{sec:seq2seq}). While we retained the original CNN encoder, we applied minimal architectural tuning and replaced the EMG cleaning steps with EEG preprocessing (Section~\ref{preprocessing}).

    \item We introduced a VAE-based EEG augmentation strategy to improve model robustness and generalization. This includes both a linear VAE and a convolutional VAE trained on \texttt{eeg\_raw} and \texttt{eeg\_feats} (see Section~\ref{sec:augvae-eeg}).
\end{enumerate}

\subsection{Proposed EEG-to-Text Model}
\label{sec:eeg2text}

\begin{figure}
    \centering
    \includegraphics[width=0.8\linewidth]{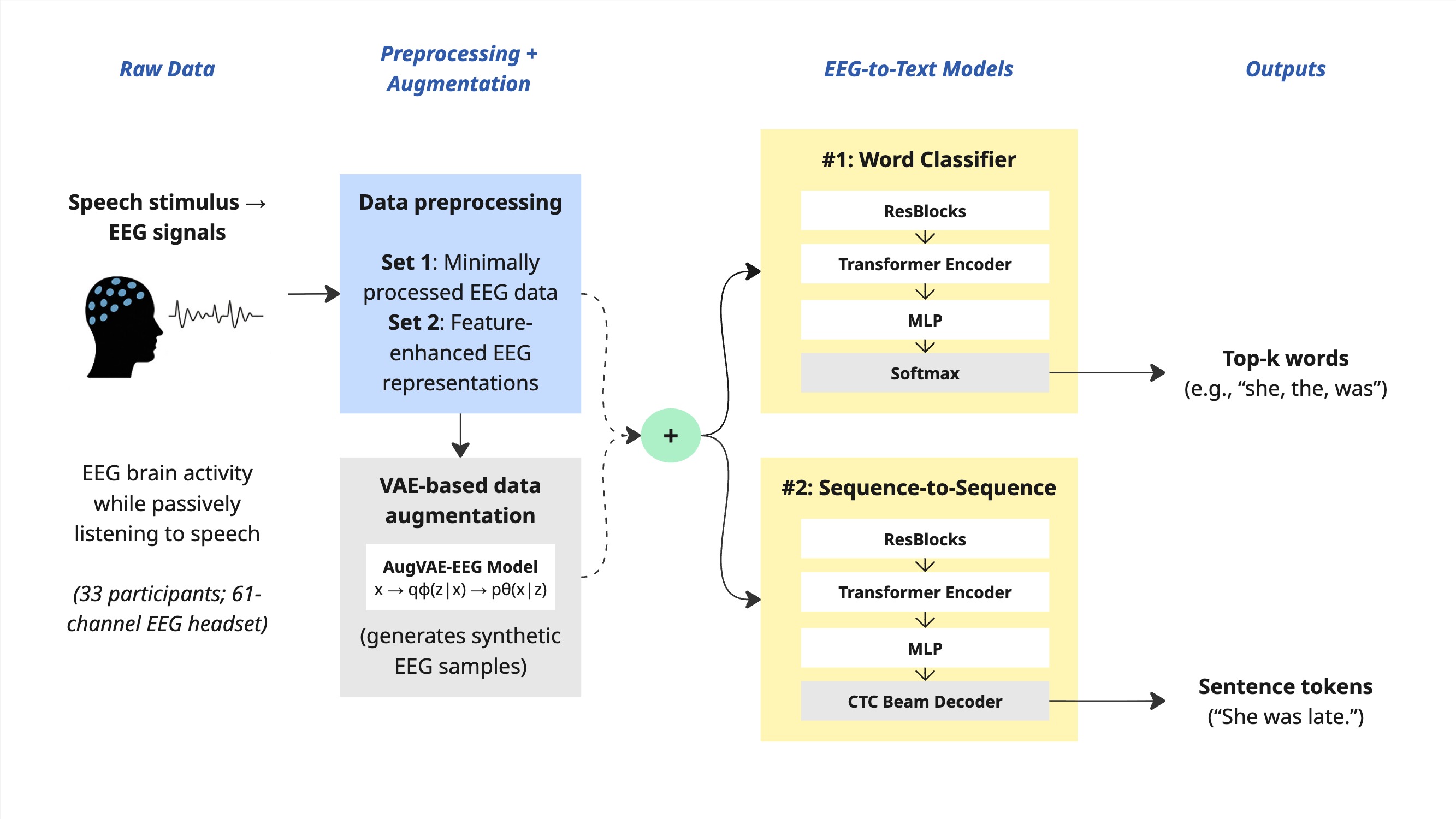}
    \caption{Approach overview.}
    \label{fig:approach}
\end{figure}

\begin{figure*}
    \centering
    \begin{subfigure}[t]{0.48\textwidth}
        \centering
        \includegraphics[width=\linewidth]{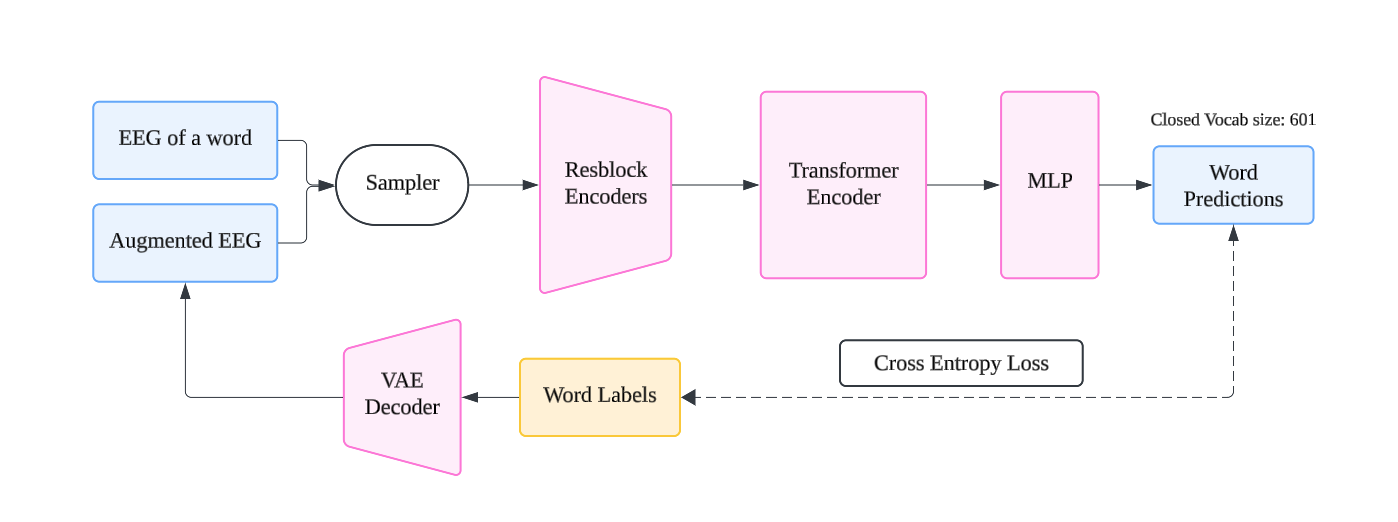}
        \caption{Word Classifier model.}
        \label{fig:eegtotext-cls-model}
    \end{subfigure}
    \hfill
    \begin{subfigure}[t]{0.48\textwidth}
        \centering
        \includegraphics[width=\linewidth]{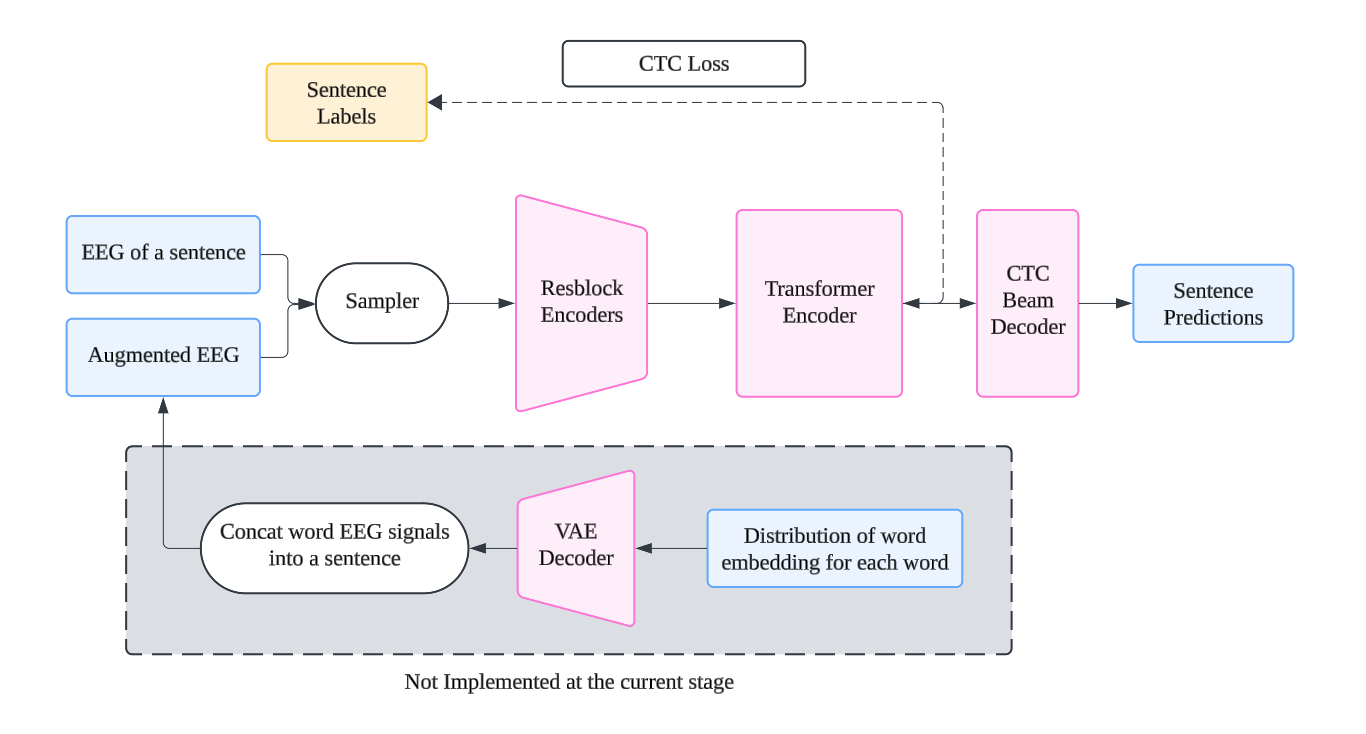}
        \caption{Sequence-to-Sequence (Seq2Seq) model.}
        \label{fig:eegtotext-seq2seq-model}
    \end{subfigure}
    \caption{EEG-to-Text models.}
    \label{fig:architecture-comparison}
\end{figure*}

We describe two architectures used for EEG decoding: a word-level classifier and a sequence-to-sequence model. See the overview of this approach on (Fig.~\ref{fig:approach}).

\subsubsection{Word Classifier Model}
\label{sec:word-classifier}
The Word Classifier model (Fig.~\ref{fig:eegtotext-cls-model}) is designed for EEG-to-word classification tasks. The corpus includes 601 unique vocabulary entries, making this a 601-class classification problem, consistent with Meta's formulation on the same dataset \cite{meta}. A full parameter summary is shown in Table~\ref{tab:wordclsmodel_summary}.

The input to the model is a preprocessed EEG segment aligned with a single word window. The architecture uses two ResBlocks to extract spatiotemporal features, followed by a Transformer encoder stack with six layers to model contextual dependencies. The final output of the transformer is passed through a linear classifier and softmax layer to produce a probability distribution over the 601-word vocabulary. The model is trained using the cross-entropy loss.

\subsubsection{Sequence-to-Sequence Model}
\label{sec:seq2seq}
The Seq2Seq model (Fig.~\ref{fig:eegtotext-seq2seq-model}) is designed for sentence-level EEG decoding and is adapted from the EMG-to-text model \cite{Gaddy_Dan_2022}. A parameter summary is presented in Table~\ref{tab:seq2seqmodel_summary}.

The model takes continuous EEG features as input, derived from preprocessing (Section~\ref{preprocessing}). Feature extraction is performed via stacked 1D convolutional ResBlocks with batch normalization and residual connections. These are followed by a linear projection to increase the representational dimension.

Temporal modeling is handled by six layers of Transformer encoders, enabling long-range attention across the EEG time series. The final output is decoded using a CTC beam search decoder, with Connectionist Temporal Classification (CTC) loss used during training to align the predicted token sequences with reference transcripts.

\begin{table*}[!t]
\centering
\begin{minipage}{0.48\textwidth}
\centering
\caption{Word Classifier model (145.68 MB)}
\label{tab:wordclsmodel_summary}
\scriptsize
\begin{tabular}{@{}lcc@{}}
\toprule
\textbf{Layer (type:depth-idx)}              & \textbf{Output Shape} & \textbf{Param \#} \\ \midrule
EEGWordClsModel                              & [8, 601]              & 768               \\
Sequential: 1-1                              & [8, 768, 130]         & --                \\
ResBlock: 2-1                                & [8, 768, 260]         & --                \\
\quad Conv1d: 3-1                            & [8, 768, 260]         & 139,008           \\
\quad BatchNorm1d: 3-2                       & [8, 768, 260]         & 1,536             \\
\quad Conv1d: 3-3                            & [8, 768, 260]         & 1,770,240         \\
\quad BatchNorm1d: 3-4                       & [8, 768, 260]         & 1,536             \\
\quad Conv1d: 3-5                            & [8, 768, 260]         & 46,848            \\
\quad BatchNorm1d: 3-6                       & [8, 768, 260]         & 1,536             \\
ResBlock: 2-2                                & [8, 768, 130]         & --                \\
\quad Conv1d: 3-7                            & [8, 768, 130]         & 1,770,240         \\
\quad BatchNorm1d: 3-8                       & [8, 768, 130]         & 1,536             \\
\quad Conv1d: 3-9                            & [8, 768, 130]         & 1,770,240         \\
\quad BatchNorm1d: 3-10                      & [8, 768, 130]         & 1,536             \\
\quad Conv1d: 3-11                           & [8, 768, 130]         & 590,592           \\
\quad BatchNorm1d: 3-12                      & [8, 768, 130]         & 1,536             \\
Linear: 1-2                                  & [8, 130, 768]         & 590,592           \\
TransformerEncoder: 1-3                      & [131, 8, 768]         & --                \\
\quad ModuleList: 2-3                        & --                    & --                \\
\quad TransformerEncoderLayer: 3-13          & [131, 8, 768]         & 7,237,632         \\
\quad TransformerEncoderLayer: 3-14          & [131, 8, 768]         & 7,237,632         \\
\quad TransformerEncoderLayer: 3-15          & [131, 8, 768]         & 7,237,632         \\
\quad TransformerEncoderLayer: 3-16          & [131, 8, 768]         & 7,237,632         \\
\quad TransformerEncoderLayer: 3-17          & [131, 8, 768]         & 7,237,632         \\
\quad TransformerEncoderLayer: 3-18          & [131, 8, 768]         & 7,237,632         \\
Linear: 1-4                                  & [8, 601]              & 462,169           \\ \bottomrule
\end{tabular}
\end{minipage}
\hfill
\begin{minipage}{0.48\textwidth}
\centering
\caption{Sequence-to-Sequence (Seq2Seq) model (143.94 MB)}
\label{tab:seq2seqmodel_summary}
\scriptsize
\begin{tabular}{@{}lcc@{}}
\toprule
\textbf{Layer (type:depth-idx)}              & \textbf{Output Shape} & \textbf{Param \#} \\ 
\midrule
EEGSeqtoSeqModel                             & [7, 1250, 38]         & --                \\
Sequential: 1-1                              & [7, 768, 1250]        & --                \\
ResBlock: 2-1                                & [7, 768, 2500]        & --                \\
\quad Conv1d: 3-1                            & [7, 768, 2500]        & 139,008           \\
\quad BatchNorm1d: 3-2                       & [7, 768, 2500]        & 1,536             \\
\quad Conv1d: 3-3                            & [7, 768, 2500]        & 1,770,240         \\
\quad BatchNorm1d: 3-4                       & [7, 768, 2500]        & 1,536             \\
\quad Conv1d: 3-5                            & [7, 768, 2500]        & 46,848            \\
\quad BatchNorm1d: 3-6                       & [7, 768, 2500]        & 1,536             \\
ResBlock: 2-2                                & [7, 768, 1250]        & --                \\
\quad Conv1d: 3-7                            & [7, 768, 1250]        & 1,770,240         \\
\quad BatchNorm1d: 3-8                       & [7, 768, 1250]        & 1,536             \\
\quad Conv1d: 3-9                            & [7, 768, 1250]        & 1,770,240         \\
\quad BatchNorm1d: 3-10                      & [7, 768, 1250]        & 1,536             \\
\quad Conv1d: 3-11                           & [7, 768, 1250]        & 590,592           \\
\quad BatchNorm1d: 3-12                      & [7, 768, 1250]        & 1,536             \\
Linear: 1-2                                  & [7, 1250, 768]        & 590,592           \\
TransformerEncoder: 1-3                      & [1250, 7, 768]        & --                \\
\quad ModuleList: 2-3                        & --                    & --                \\
\quad TransformerEncoderLayer: 3-13          & [1250, 7, 768]        & 7,237,632         \\
\quad TransformerEncoderLayer: 3-14          & [1250, 7, 768]        & 7,237,632         \\
\quad TransformerEncoderLayer: 3-15          & [1250, 7, 768]        & 7,237,632         \\
\quad TransformerEncoderLayer: 3-16          & [1250, 7, 768]        & 7,237,632         \\
\quad TransformerEncoderLayer: 3-17          & [1250, 7, 768]        & 7,237,632         \\
\quad TransformerEncoderLayer: 3-18          & [1250, 7, 768]        & 7,237,632         \\
Linear: 1-4                                  & [7, 1250, 38]         & 29,222            \\ 
\bottomrule
\end{tabular}
\end{minipage}
\end{table*}

\subsection{Proposed AugVAE-EEG Model}
\label{sec:augvae-eeg}

Variational Autoencoders (VAEs) provide a principled framework for learning compressed, noise-resilient representations of high-dimensional data, making them particularly suitable for EEG augmentation. A VAE consists of an encoder $\mathbf{q}_\phi(\mathbf{z}|\mathbf{x})$ that maps input data $\mathbf{x}$ to a latent distribution $\mathbf{z}$, and a decoder $p_\theta(\mathbf{x}|\mathbf{z})$ that reconstructs the input from latent samples. The model is trained using the Evidence Lower Bound (ELBO) objective, which balances reconstruction fidelity and regularization of the latent space.

\begin{equation}
\begin{split}
L(\mathbf{\theta}; \mathbf{\phi} ; \mathbf{x}^{(i)}) = 
& \ \frac{1}{2} \sum_{j=1}^{J} \Big( 1 + \log((\sigma^{(i)}_j)^2) - (\mu^{(i)}_j)^2 - (\sigma^{(i)}_j)^2 \Big) \\
& + \frac{1}{L} \sum_{l=1}^{L} \log \mathbf{p_\theta} \big( \mathbf{x}^{(i)} \mid \mathbf{z}^{(i,l)} \big)
\end{split}
\label{eq:vae_loss}
\end{equation}

To allow gradient-based optimization, the reparameterization trick is applied as:
\begin{equation}
\mathbf{z}^{(i)} = \mu^{(i)} + \sigma^{(i)} \cdot \boldsymbol{\epsilon}^{(i)}, 
\quad \boldsymbol{\epsilon}^{(i)} \sim \mathcal{N}(0, I)
\label{eq:reparam}
\end{equation}

VAEs have been successfully used for generating synthetic data to augment limited datasets across various domains. Prior work has demonstrated their effectiveness in boosting classification performance in speech and medical tasks via MLP-VAE, CNN-VAE, and conditional variants \cite{saldanha2022data, 8282225}.

In this study, we design a lightweight VAE architecture for EEG data. The encoder and decoder each consist of two fully connected layers with sizes 512 and 256, activated by ReLU. The latent space has a dimension of 64. The VAE is trained on \texttt{eeg\_raw} signals (raw EEG windows) from 10 subjects with high comprehension scores. Although models were also trained on \texttt{eeg\_feats} (manually extracted features described in Section~\ref{preprocessing}), we find that \texttt{eeg\_raw} leads to better generation quality and is used in the final pipeline.

After VAE training, we compute the mean and standard deviation of latent vectors for each word across subjects. During training of the downstream decoder, with a certain probability, we replace (or supplement) real samples with synthetic EEG. These are generated by sampling from the learned Gaussian latent space and decoding to the EEG domain. This allows the model to encounter diverse but realistic examples of the same label, improving generalization.

\begin{figure}[!t]
    \centering
    \includegraphics[width=0.8\linewidth]{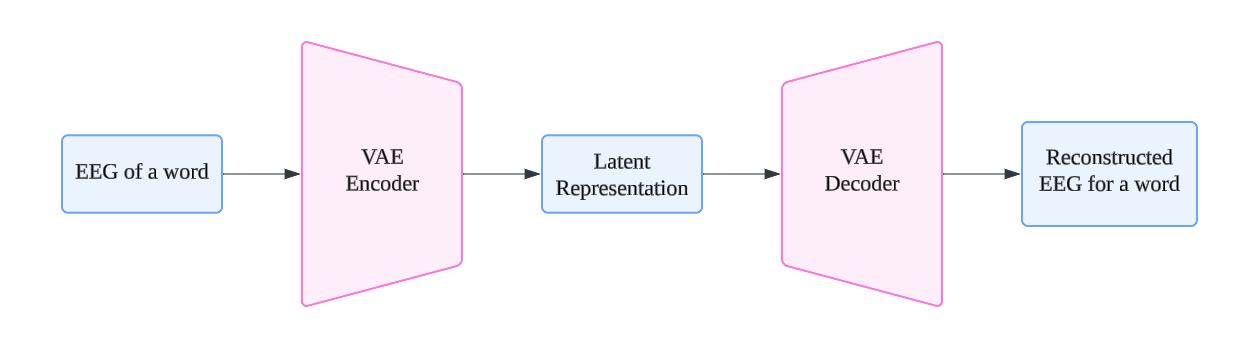}
    \caption{AugVAE-EEG model.}
    \label{fig:AUGVAE}
\end{figure}

\begin{figure}[!t]
    \centering
    \includegraphics[width=0.7\linewidth]{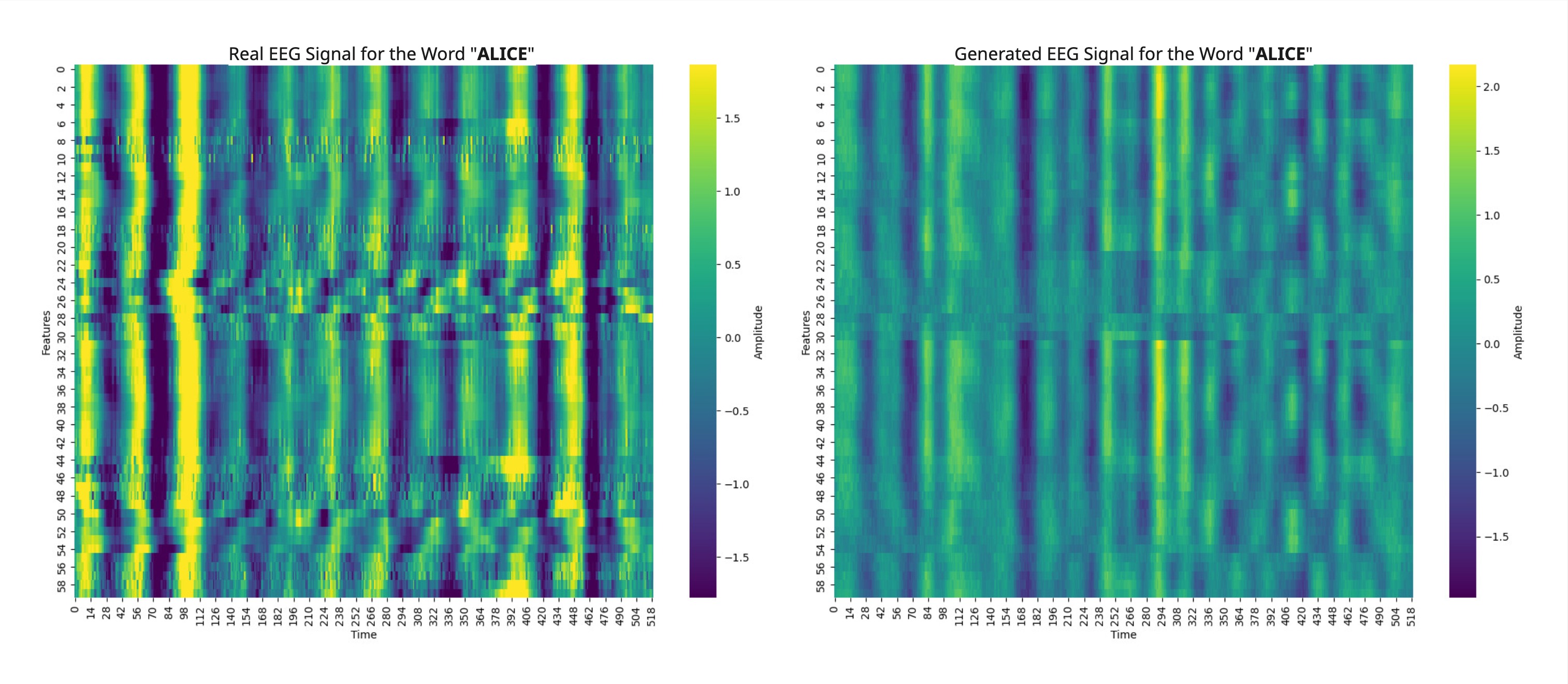}
    \caption{Comparison of real and generated EEG signals.}
    \label{augmentation_example}
\end{figure}

\section{Evaluation Metrics}

We used different evaluation metrics for the two EEG-to-text models, based on the nature of their outputs.

For the Word Classifier model, we used \textit{accuracy} as the primary evaluation metric. Accuracy measures the proportion of correctly classified samples relative to the total number of samples. It is defined as:
\begin{equation}
\text{Accuracy} = \frac{C}{N},
\end{equation}
where $C$ is the number of samples correctly classified by the model, and $N$ is the total number of samples in the dataset.

For the Seq2Seq model, we used the \textit{Word Error Rate (WER)}, which assesses the intelligibility of the generated output by comparing it to the reference text. WER is calculated as:
\begin{equation}
\text{WER} = \frac{S + I + D}{R},
\end{equation}
where $S$ is the number of substitutions, $I$ is the number of insertions, $D$ is the number of deletions, and $R$ is the total number of words in the reference text. 

\section{Loss Functions}

The choice of loss function is critical to model performance, as it determines how the model learns from data. In this study, we use different loss functions for the classification, sequence modeling, and generative components of our system.

\subsection{Cross-Entropy Loss (Word Classifier)}

For the Word Classifier model, we use the standard cross-entropy loss, commonly applied in multi-class classification tasks. It measures the divergence between the true label distribution and the predicted probability distribution. The loss is defined as:
\begin{equation}
\mathcal{L}_{\text{CE}} = -\frac{1}{N} \sum_{i=1}^{N} \sum_{j=1}^{C} y_{ij} \log \hat{y}_{ij},
\end{equation}
where $N$ is the number of samples, $C$ is the number of classes (601 in our case), $y_{ij}$ is the binary indicator (0 or 1) of whether class $j$ is the correct class for sample $i$, and $\hat{y}_{ij}$ is the predicted probability for class $j$.

\subsection{Connectionist Temporal Classification (CTC) Loss (Seq2Seq)}

The Seq2Seq model is trained using Connectionist Temporal Classification (CTC) loss, which is designed for sequence prediction tasks with unaligned input-output pairs. CTC computes the negative log-likelihood of the correct output sequence $\mathbf{y}$ given the input sequence $\mathbf{x}$:
\begin{equation}
\mathcal{L}_{\text{CTC}} = -\log P(\mathbf{y} \mid \mathbf{x}).
\end{equation}

This probability is computed by summing over all valid alignments between input and output sequences, allowing the model to learn both the token sequence and its timing implicitly. CTC is particularly suitable when input and output lengths differ, as is common in speech and EEG decoding.

\subsection{Variational Autoencoder (VAE) Loss (AugVAE-EEG)}

The loss function for the AugVAE-EEG model combines two objectives: reconstructing the input EEG signal from its latent representation, and regularizing the latent space to match a prior distribution.

\begin{equation}
\mathcal{L}_{\text{VAE}} = \mathcal{L}_{\text{reconstruction}} + \beta \cdot \mathcal{L}_{\text{KL}},
\end{equation}

Here, $\mathcal{L}_{\text{reconstruction}}$ ensures the decoder output closely resembles the input, while $\mathcal{L}_{\text{KL}}$ encourages the encoded latent variables to follow a standard normal distribution. The hyperparameter $\beta$ controls the trade-off between the two terms.

\subsubsection*{Reconstruction Loss}

To quantify the reconstruction error, we use Mean Squared Error (MSE) between the original input $x$ and the reconstructed signal $\hat{x}$:
\begin{equation}
\mathcal{L}_{\text{reconstruction}}(x, \hat{x}) = \|x - \hat{x}\|_2^2.
\end{equation}

\subsubsection*{KL Divergence Loss}

The KL divergence term measures how much the learned latent distribution $q(z|x)$ deviates from the prior $p(z)$. Assuming a standard normal prior, the closed-form KL divergence is:
\begin{equation}
\mathcal{L}_{\text{KL}}(\mu, \sigma^2) = \frac{1}{2} \sum_{j=1}^{d} \left( 1 + \log(\sigma^2_j) - \mu_j^2 - \sigma^2_j \right),
\end{equation}
where $\mu$ and $\sigma^2$ are the mean and variance of the latent space for each dimension $j = 1, \dots, d$.

\section{Results} 

\subsection{Baseline Replication: EMG-to-Speech}
We replicated the EMG-to-speech model from \cite{Gaddy_Dan_2022} using the original public repository. Our implementation achieved a word error rate (WER) of 34.9\% on the open dataset, closely matching the reported 36.1\% WER. Due to this alignment, no further optimization was performed. For reproducibility and ease of access, we provide an executable notebook version at \url{https://github.com/YHTerrance/silent_speech/blob/main/GaddyNB.ipynb}.

\subsection{EEG-to-Text Experiments}
We evaluated two architectures adapted for EEG decoding: a \textbf{Word Classifier} and a \textbf{Sequence-to-Sequence (Seq2Seq)} model. We conducted a series of ablation studies using techniques such as VAE-based data augmentation, stratified sampling, and input masking. Results are reported in terms of top-1/top-10 word accuracy for the classifier and WER for the Seq2Seq model.

\subsubsection{Word Classifier Model}
\begin{figure}[!t]
    \centering
    \begin{subfigure}[b]{0.4\linewidth}
        \centering
        \includegraphics[width=\linewidth]{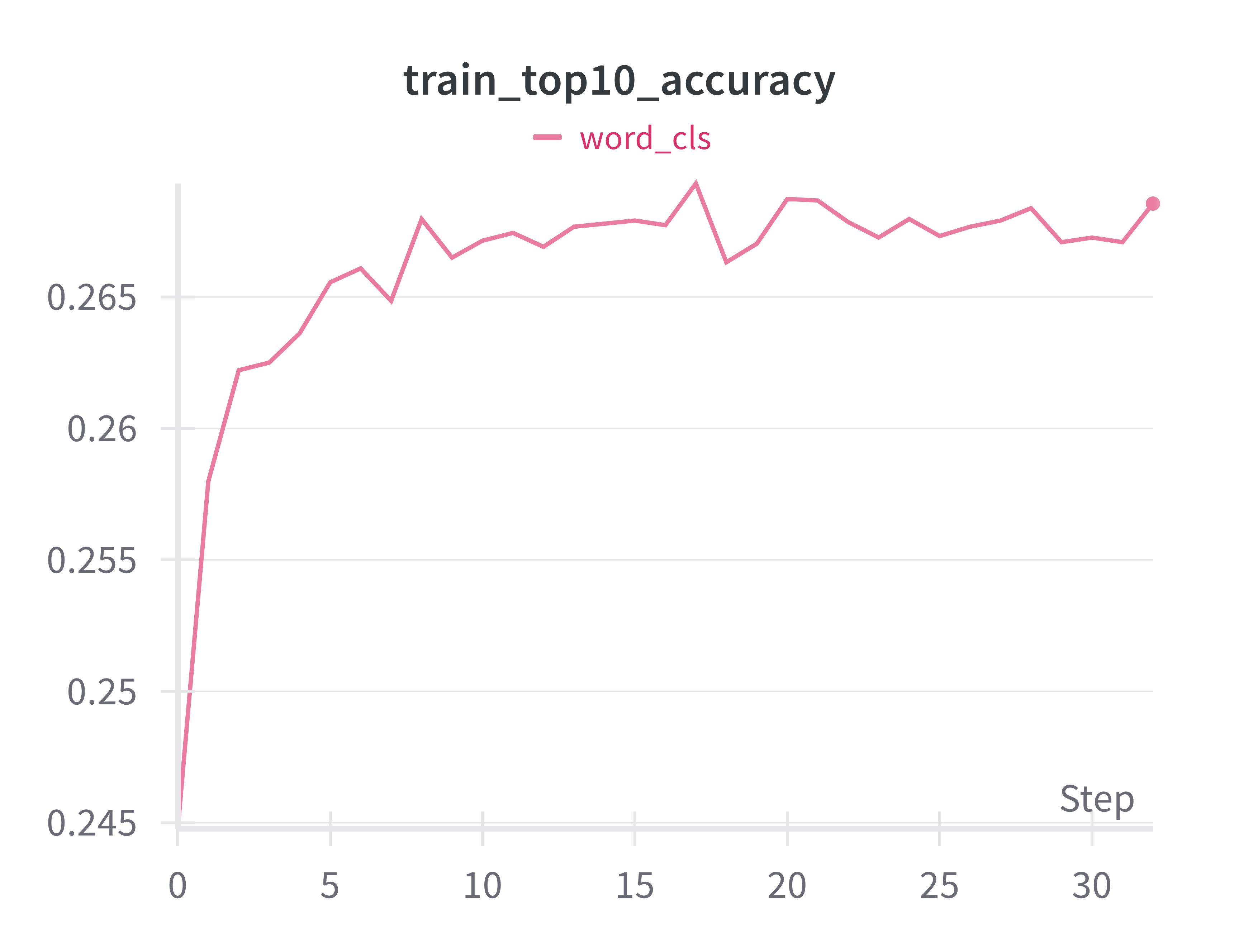}
        \caption{Training Top-10 Accuracy}
        \label{fig:word_cls_train_acc}
    \end{subfigure}
    \hfill
    \begin{subfigure}[b]{0.4\linewidth}
        \centering
        \includegraphics[width=\linewidth]{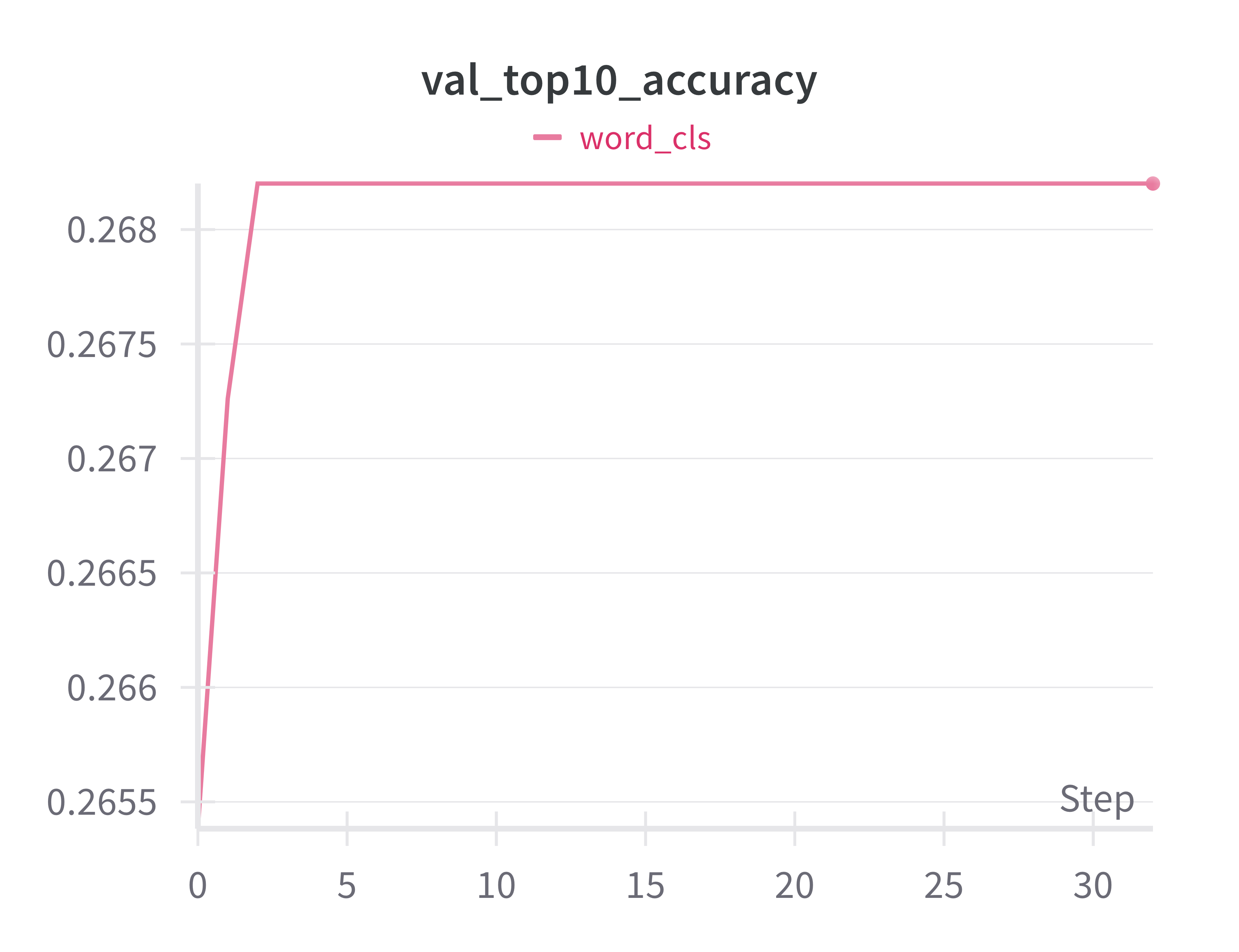}
        \caption{Validation Top-10 Accuracy}
        \label{fig:word_cls_val_acc}
    \end{subfigure}
    \caption{Top-10 accuracy of the Word Classifier model.}
    \label{fig:word_cls_accuracy}
\end{figure}

The classifier predicts individual words from EEG across 601 unique classes. In the baseline setting using stratified sampling and masking, it achieved a Top-1 accuracy of 4.1\% and a Top-10 accuracy of 26.82\% on the validation set, consistent with Meta’s benchmark of 25.7\% Top-10 accuracy on the same dataset \cite{meta}. However, predictions were skewed toward high-frequency tokens (e.g., ``she'', ``the'', ``was'', ``it'', ``and''), suggesting the model exploited word distribution priors rather than learning EEG-specific features. See Fig.~\ref{fig:word_cls_train_acc} and Fig.~\ref{fig:word_cls_val_acc} for training and validation curves.

To encourage rare word learning, we tried inverse word frequency weighting in the loss. This led to a drastic performance drop, with Top-10 accuracy falling below 0.1\%. This highlights the challenge of meaningful word prediction from EEG.

\subsubsection{Sequence-to-Sequence Model}
\begin{figure}[!t]
    \centering
    \begin{subfigure}[b]{0.4\linewidth}
        \centering
        \includegraphics[width=\linewidth]{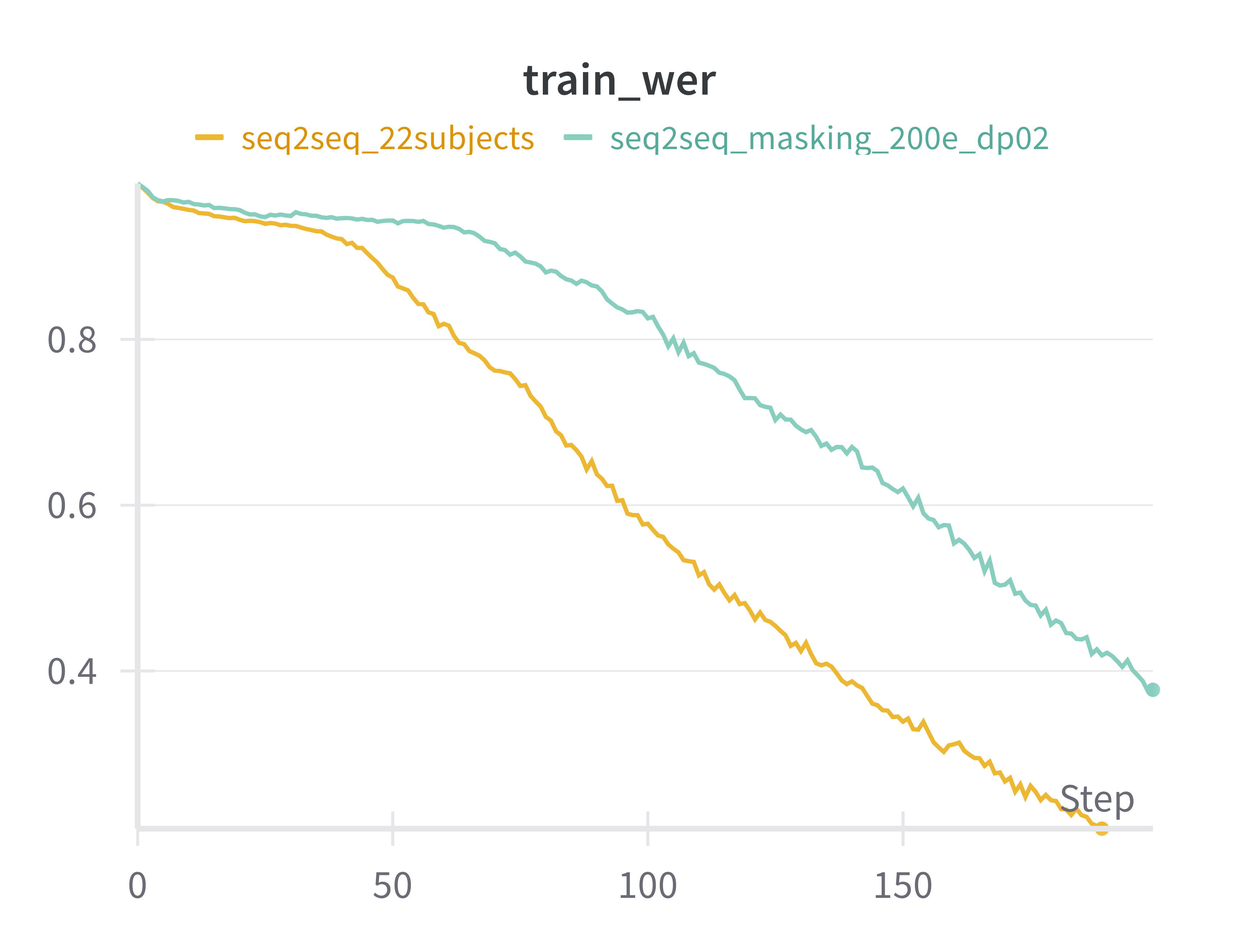}
        \caption{Training WER}
        \label{fig:seq2seq_train_wer}
    \end{subfigure}
    \hfill
    \begin{subfigure}[b]{0.4\linewidth}
        \centering
        \includegraphics[width=\linewidth]{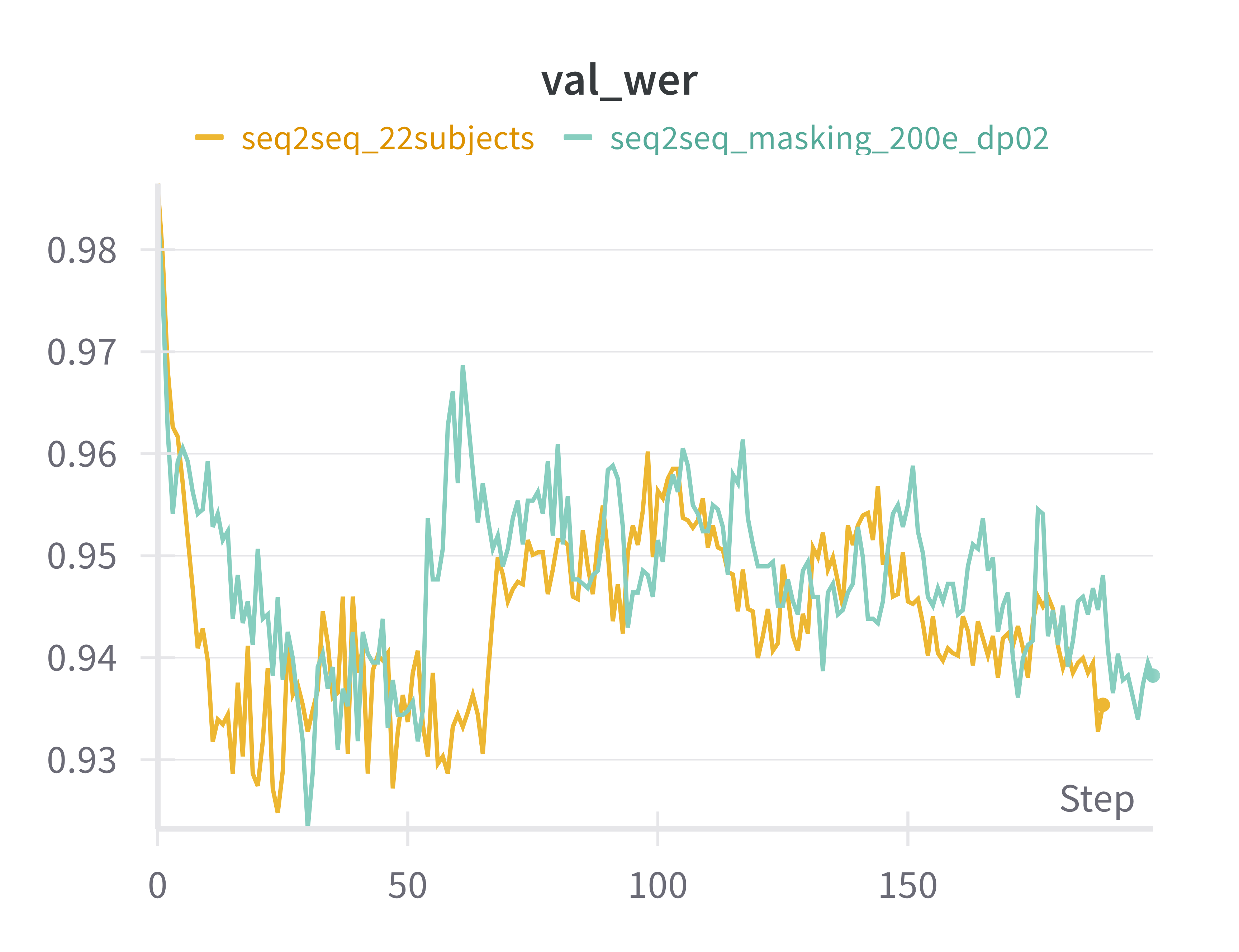}
        \caption{Validation WER}
        \label{fig:seq2seq_val_wer}
    \end{subfigure}
    \caption{WER with and without masking over 200 epochs for the Seq2Seq model.}
    \label{fig:seq2seq_masking}
\end{figure}

The Seq2Seq model aimed to decode full sentences from EEG inputs. In initial training across 22 subjects, the model overfit severely: training WER reached 20.97\%, while validation WER plateaued at 92.48\%. Adding time and frequency masking marginally improved generalization (validation WER: 92.32\%; see Table~\ref{tab:seq2seq_results} and Fig.~\ref{fig:seq2seq_train_wer}, Fig.~\ref{fig:seq2seq_val_wer}).

\begin{figure}[!t]
    \centering
    \begin{subfigure}[b]{0.4\linewidth}
        \centering
        \includegraphics[width=\linewidth]{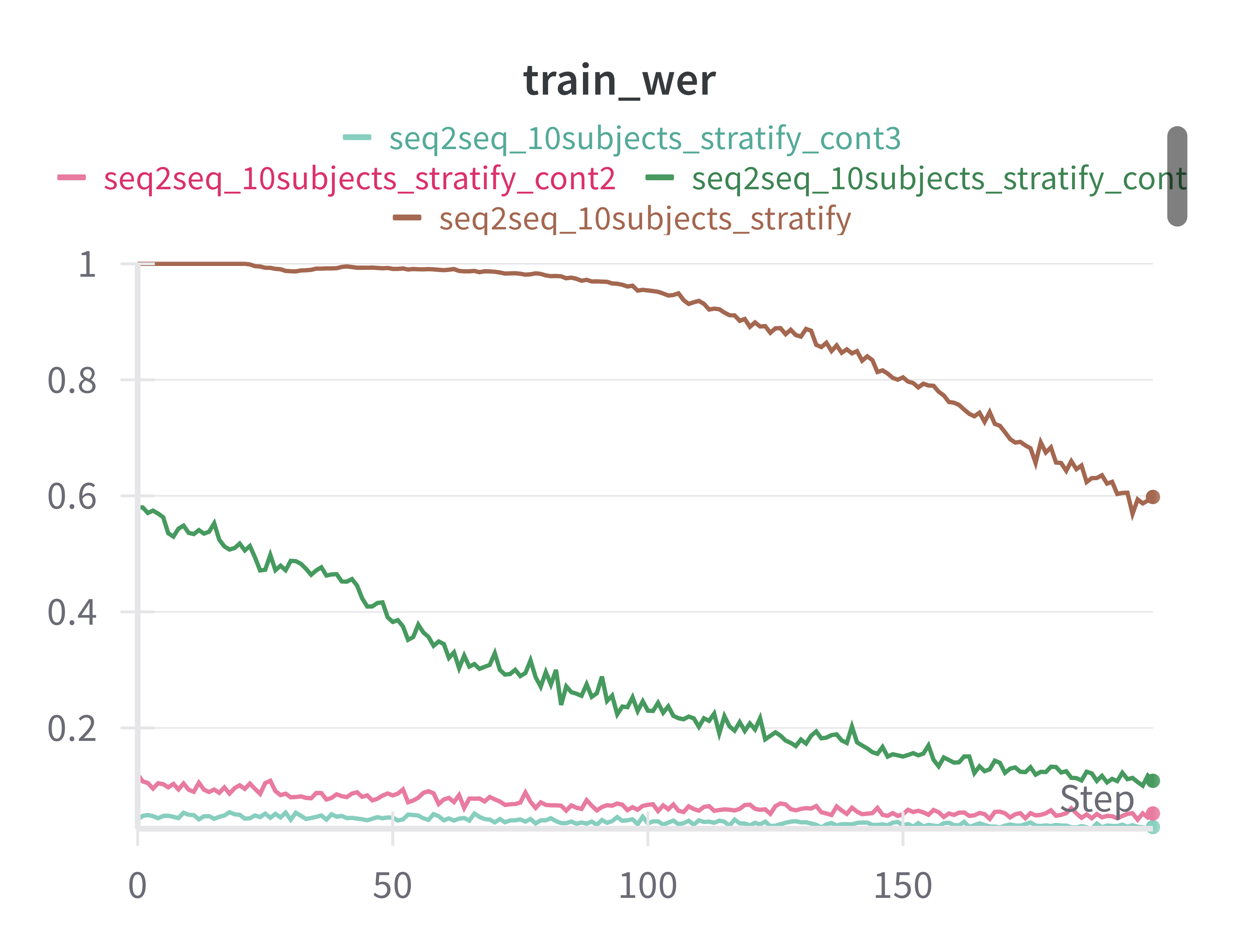}
        \caption{Training WER}
        \label{fig:strat_train_wer}
    \end{subfigure}
    \hfill
    \begin{subfigure}[b]{0.4\linewidth}
        \centering
        \includegraphics[width=\linewidth]{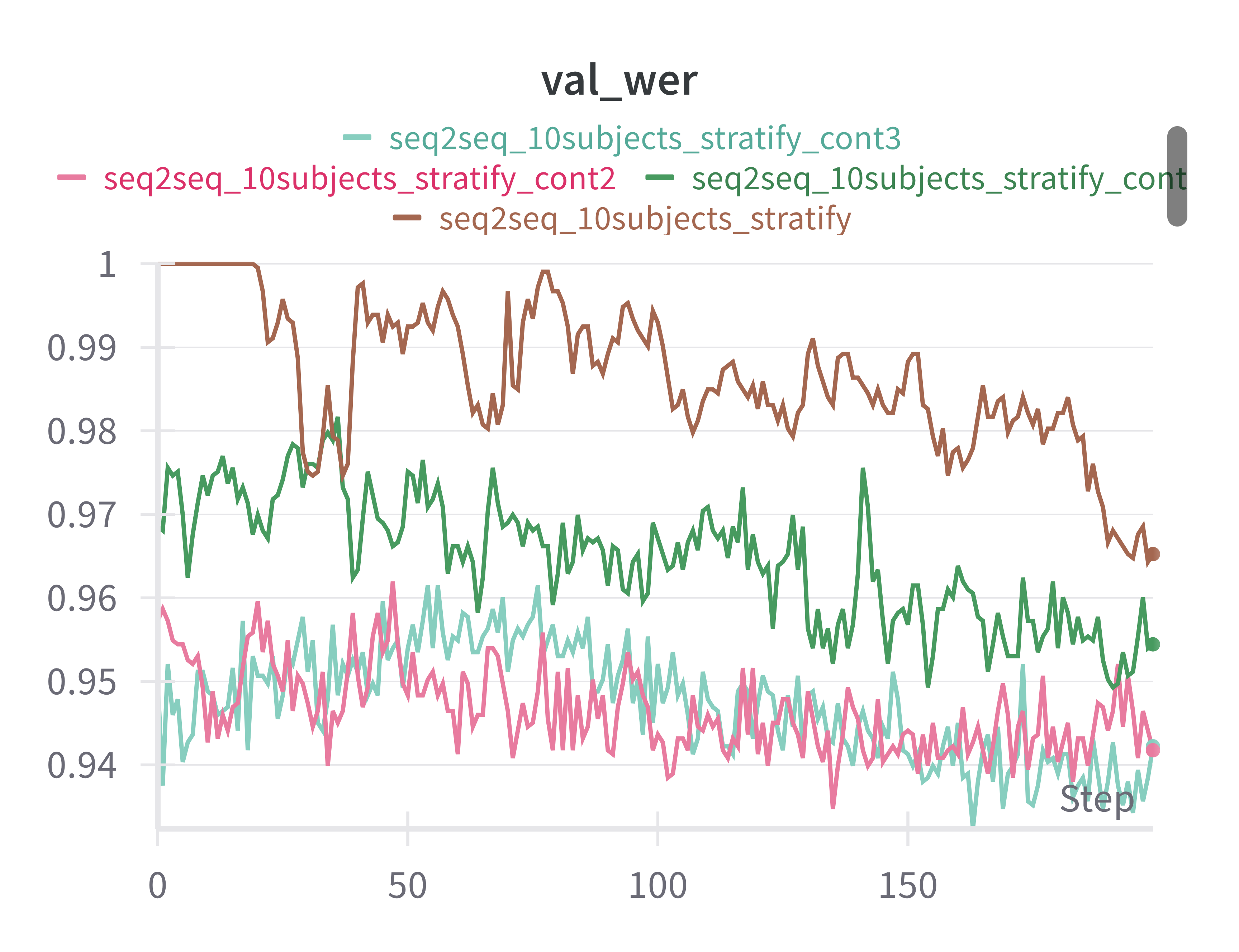}
        \caption{Validation WER}
        \label{fig:strat_val_wer}
    \end{subfigure}
    \caption{WER with stratified sampling over 800 epochs for the Seq2Seq model.}
    \label{fig:seq2seq_stratified}
\end{figure}

We discovered data leakage due to overlapping sentence distributions across training and validation sets. A sentence-level stratified sampling strategy was introduced and evaluated on 10 subjects. The model trained for 800 epochs: training WER dropped to 2.91\%, while validation WER stabilized at 93.23\% after 400 epochs (Fig.~\ref{fig:strat_train_wer}, Fig.~\ref{fig:strat_val_wer}). Despite mitigation strategies, generalization remained poor.

\begin{table}[!t]
    \centering
    \scriptsize
    \caption{Performance Metrics for the Seq2Seq Model.}
    \label{tab:seq2seq_results}
    \begin{tabular}{@{\hskip 1pt}l@{\hskip 4pt}c@{\hskip 4pt}c@{\hskip 4pt}c@{\hskip 4pt}c@{\hskip 1pt}}
        \toprule
        \textbf{Experiment} & \textbf{Epochs} & \textbf{Train Loss} & \textbf{Train WER (\%)} & \textbf{Val WER (\%)} \\
        \midrule
        22 subjects & 200 & 0.35 & 20.97 & 92.48 \\
        + Masking & 200 & 0.50 & 37.73 & 92.32 \\
        10 subj. + Strat. + Masking & 800 & 0.06 & 2.91 & 93.23 \\
        \bottomrule
    \end{tabular}
\end{table}

\subsection{AugVAE-EEG: VAE-based Data Augmentation}
\begin{figure}[!t]
    \centering
    \begin{subfigure}[b]{0.47\linewidth}
        \centering
        \includegraphics[width=\linewidth]{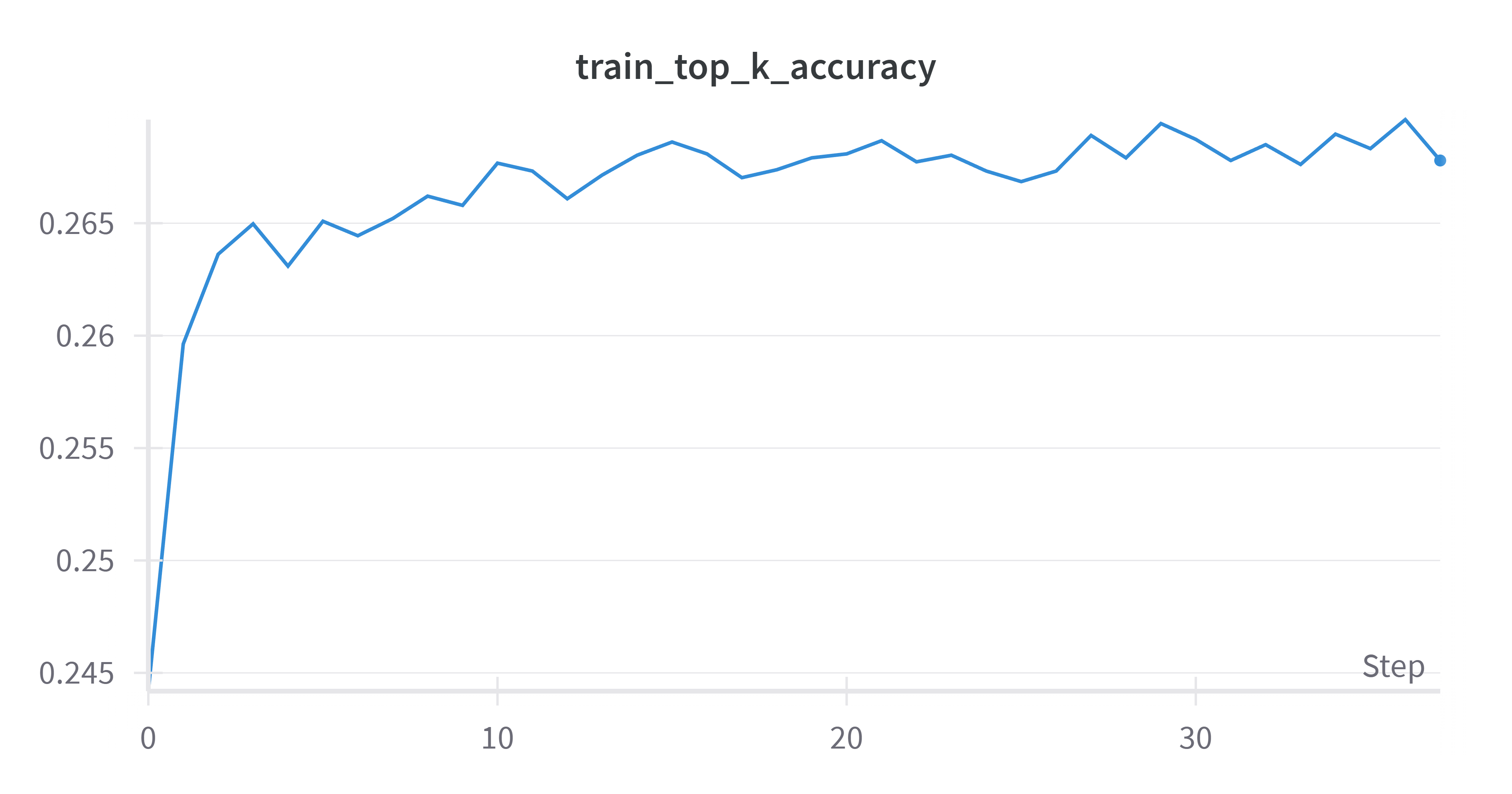}
        \caption{Training Top-10 Accuracy}
        \label{fig:vae_train_acc}
    \end{subfigure}
    \hfill
    \begin{subfigure}[b]{0.47\linewidth}
        \centering
        \includegraphics[width=\linewidth]{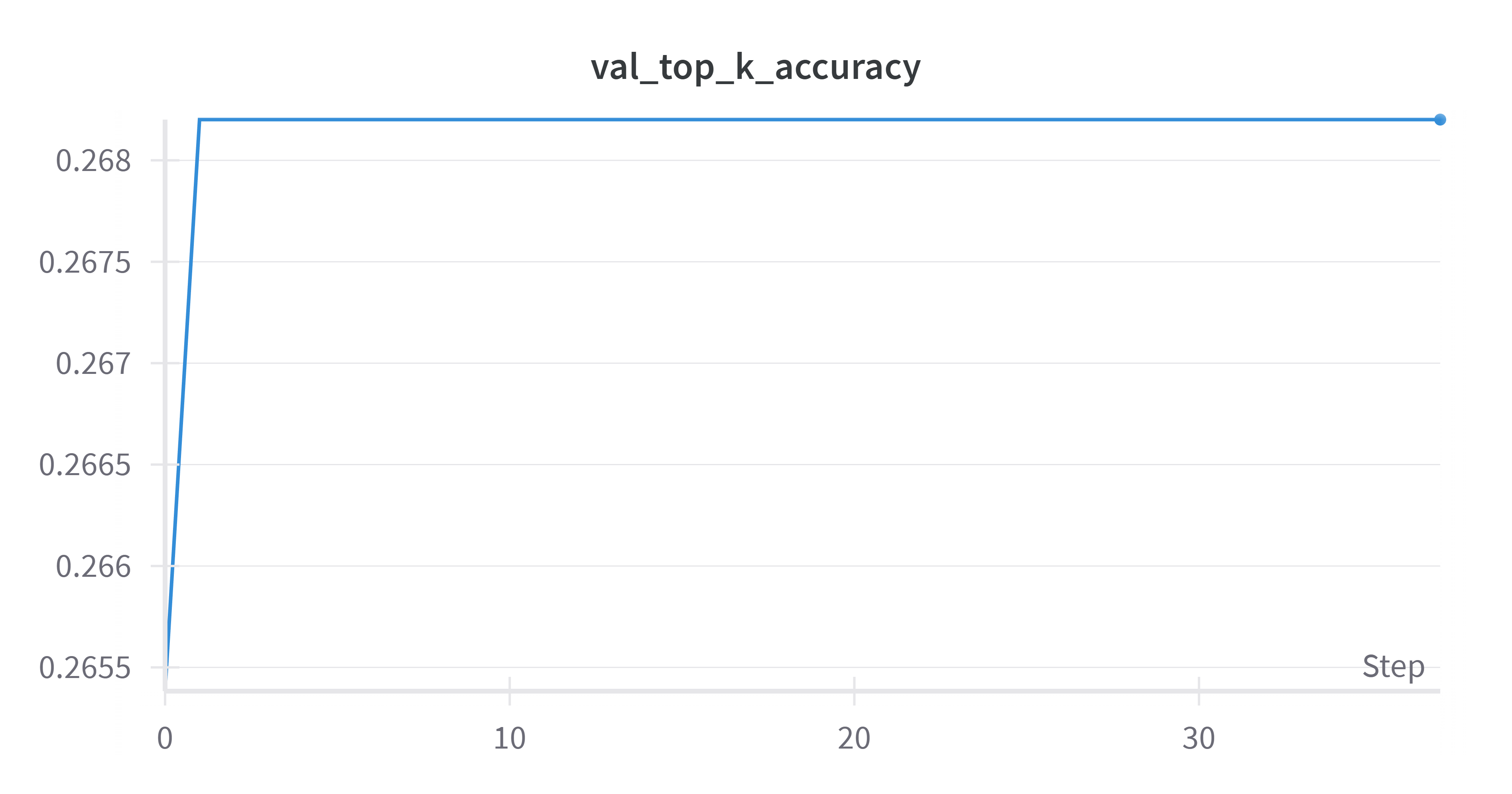}
        \caption{Validation Top-10 Accuracy}
        \label{fig:vae_val_acc}
    \end{subfigure}
    \caption{Top-10 accuracy for the Word Classifier model with 50\% VAE-augmented EEG signals.}
    \label{fig:vae_aug_results}
\end{figure}

We evaluated whether augmenting EEG inputs with VAE-generated signals could improve classifier generalization. Replacing 50\% of input EEG samples with VAE reconstructions yielded no performance gain: top-10 accuracy converged to the same level as the baseline (Fig.~\ref{fig:vae_train_acc}, Fig.~\ref{fig:vae_val_acc}). Increasing augmentation to 90\% produced similar results. These findings suggest that the VAE-generated data did not inject sufficient diversity or informativeness to aid model learning.

\section{Discussion}
Our results highlight both the promise and limitations of EEG-based speech decoding. The Word Classifier achieved a Top-10 accuracy of 26.82\%, consistent with prior work \cite{meta}, yet its predictions were dominated by high-frequency words. The Seq2Seq model captured temporal structure during training (WER: 2.91\%) but failed to generalize (val WER: 93.23\%). VAE-based augmentation and masking techniques provided negligible gains, pointing to deeper limitations in data or modeling. Key observations are the following:

\begin{itemize}
    \item \textbf{Masking effects:} Time and frequency masking slightly reduced validation WER (92.48\% to 92.32\%) but failed to prevent overfitting. This suggests masking alone is insufficient for regularization in low-signal EEG regimes.

    \item \textbf{Ineffectiveness of VAE augmentation:} Augmenting 50-90\% of inputs with VAE-generated EEG showed no improvement. The synthetic signals may not introduce meaningful diversity or realism.

    \item \textbf{Stratified sampling:} While stratification stabilized training, it did not improve validation WER. Sentence imbalance alone is not the root cause of generalization failure.

    \item \textbf{Model comparison:} The Word Classifier is sensitive to word frequency, lacking deeper signal understanding. The Seq2Seq model, though temporally expressive, lacks robustness across validation subjects.

    \item \textbf{Loss weighting:} Inverse frequency weighting collapsed performance (Top-10 accuracy $\sim$0.1\%), confirming that the classifier heavily exploits frequency priors and fails to learn general EEG-word associations.
\end{itemize}

\noindent Overall, these findings suggest that decoding text from EEG requires either stronger architectural priors or more diverse, subject-agnostic training data. Future work should explore cross-subject normalization, large-scale pretraining, and hybrid contrastive losses.

\section{Future Work}
Models explored in this project show promise for future assistive technologies. One direction is to extend our AugVAE-EEG pipeline to the sequence-to-sequence task, potentially improving generalization by introducing more diverse training signals. Another is to apply the approach to silent or imagined speech decoding, which is more applicable to real-world use cases, though hindered by a lack of public datasets. Finally, incorporating additional modalities (e.g., audio or phonemes) may enable richer, multimodal representations and improve decoding performance.

\section{Conclusions}
Our results highlight the challenges of EEG-to-text decoding, with models struggling to generalize despite capturing some temporal structure. While VAEs offered a starting point for augmentation, they require further development to improve robustness. Still, the Seq2Seq model's ability to learn sentence dynamics suggests that a meaningful signal exists in the data. This work establishes a baseline and identifies key limitations, paving the way for future research on more powerful architectures, better augmentation, and multimodal learning for practical brain-to-text systems.

\section{Code}
All code used for training is available on our GitHub repository, forked from \cite{gaddy2020digital} and adapted for EEG-based speech decoding: \url{https://github.com/YHTerrance/silent_speech}. Logs from our ablation studies are accessible at \url{https://wandb.ai/cmu-idl-best/eeg-alice?nw=nwuserjolinc}.

\section*{Acknowledgments}

The authors would like to thank Professor Bhiksha Raj of Carnegie Mellon University for his guidance and support.

\clearpage
\bibliographystyle{abbrvnat}
\bibliography{references}

@article{Shamlo2015The,title={The PREP pipeline: standardized preprocessing for large-scale EEG analysis},author={Nima Bigdely Shamlo and T. Mullen and Christian Kothe and Kyungmin Su and K. Robbins},journal={Frontiers in Neuroinformatics},year={2015},volume={9},doi={10.3389/fninf.2015.00016}}

@article{gaddy2020digital,
  author    = {Gaddy, David and Klein, Dan},
  title     = {Digital Voicing of Silent Speech},
  year      = {2020},
  journal   = {arXiv preprint arXiv:2010.02960},
  note      = {EMNLP 2020},
  doi       = {10.48550/arXiv.2010.02960},
  subjects   = {Audio and Speech Processing (eess.AS); Computation and Language (cs.CL); Machine Learning (cs.LG); Sound (cs.SD)}
}

@article{LSTMRNN_EEGClassification,
  author = {Abdulghani, M.M. and Walters, W.L. and Abed, K.H.},
  title = {Classification Using EEG and Deep
  Learning},
  journal = {Bioengineering 2023},
  volume = {10},
  pages = {649},
  doi       = {10.3390/},
  received  = {5 May 2023},
  accepted  = {25 May 2023},
  published  = {26 May 2023},
  year = {2023}
}

@article{Nguyen_2018,
doi = {10.1088/1741-2552/aa8235},
url = {https://dx.doi.org/10.1088/1741-2552/aa8235},
year = {2017},
month = {dec},
publisher = {IOP Publishing},
volume = {15},
number = {1},
pages = {016002},
author = {Chuong H Nguyen and George K Karavas and Panagiotis Artemiadis},
title = {Inferring imagined speech using EEG signals: a new approach using Riemannian manifold features},
journal = {Journal of Neural Engineering}
}

@inproceedings{German_etal_2017,
author = {Germ{\'a}n A. Pressel Coretto and Iv{\'a}n E. Gareis and H. Leonardo Rufiner},
title = {{Open access database of EEG signals recorded during imagined speech}},
volume = {10160},
booktitle = {12th International Symposium on Medical Information Processing and Analysis},
editor = {Eduardo Romero and Natasha Lepore and Jorge Brieva and Jorge Brieva and Ignacio Larrabide and },
organization = {International Society for Optics and Photonics},
publisher = {SPIE},
pages = {1016002},
year = {2017},
doi = {10.1117/12.2255697},
URL = {https://doi.org/10.1117/12.2255697}
}

@article{crossmodal,
  title         = {A Cross-Modal Approach to Silent Speech with LLM-Enhanced Recognition},
  author        = {Tyler Benster and Guy Wilson and Reshef Elisha and Francis R. Willett and Shaul Druckmann},
  journal       = {arXiv preprint arXiv:2403.05583},
  year          = {2024},
  eprint        = {2403.05583},
  archivePrefix = {arXiv},
  primaryClass  = {cs.HC},
  url           = {https://arxiv.org/abs/2403.05583}
}

@article{DAS2024,
title = {Multimodal speech recognition using EEG and audio signals: A novel approach for enhancing ASR systems},
journal = {Smart Health},
volume = {32},
pages = {100477},
year = {2024},
issn = {2352-6483},
doi = {https://doi.org/10.1016/j.smhl.2024.100477},
url = {https://www.sciencedirect.com/science/article/pii/S2352648324000333},
author = {Anarghya Das and Puru Soni and Ming-Chun Huang and Feng Lin and Wenyao Xu}
}

@article{highperformance_speech,
  author    = {Willett, Francis R. and Kunz, Erin M. and Fan, Chaofei and Avansino, Donald T. and Wilson, Guy H. and Choi, Eun Young and Kamdar, Foram and Glasser, Matthew F. and Hochberg, Leigh R. and Druckmann, Shaul and Shenoy, Krishna V. and Henderson, Jaimie M.},
  title     = {Thinking out loud, an open-access EEG-based BCI dataset for inner speech recognition},
  journal   = {Scientific Data},
  volume    = {9},
  pages     = {52},
  year      = {2022},
  doi       = {10.1038/s41597-022-01147-2},
  received  = {20 April 2021},
  accepted  = {23 December 2021},
  published  = {14 February 2022}
}

@article{DENBY2010270,
title = {Silent speech interfaces},
journal = {Speech Communication},
volume = {52},
number = {4},
pages = {270-287},
year = {2010},
note = {Silent Speech Interfaces},
issn = {0167-6393},
doi = {https://doi.org/10.1016/j.specom.2009.08.002},
url = {https://www.sciencedirect.com/science/article/pii/S0167639309001307},
author = {B. Denby and T. Schultz and K. Honda and T. Hueber and J.M. Gilbert and J.S. Brumberg}
}

@article{meta,
   title={Decoding speech perception from non-invasive brain recordings},
   volume={5},
   ISSN={2522-5839},
   url={http://dx.doi.org/10.1038/s42256-023-00714-5},
   DOI={10.1038/s42256-023-00714-5},
   number={10},
   journal={Nature Machine Intelligence},
   publisher={Springer Science and Business Media LLC},
   author={Défossez, Alexandre and Caucheteux, Charlotte and Rapin, Jérémy and Kabeli, Ori and King, Jean-Rémi},
   year={2023},
   month=oct, pages={1097–1107} }

@article{CAI2024106131,
title = {MAE-EEG-Transformer: A transformer-based approach combining masked autoencoder and cross-individual data augmentation pre-training for EEG classification},
journal = {Biomedical Signal Processing and Control},
volume = {94},
pages = {106131},
year = {2024},
issn = {1746-8094},
doi = {https://doi.org/10.1016/j.bspc.2024.106131},
url = {https://www.sciencedirect.com/science/article/pii/S1746809424001897},
author = {Miao Cai and Yu Zeng}
}

@misc{chien2022maeegmaskedautoencodereeg,
      title={MAEEG: Masked Auto-encoder for EEG Representation Learning}, 
      author={Hsiang-Yun Sherry Chien and Hanlin Goh and Christopher M. Sandino and Joseph Y. Cheng},
      year={2022},
      eprint={2211.02625},
      archivePrefix={arXiv},
      primaryClass={eess.SP},
      url={https://arxiv.org/abs/2211.02625}, 
}

@phdthesis{Gaddy_Dan_2022, title={Voicing silent speech}, journal={Voicing Silent Speech}, school={eScholarship, University of California, Berkeley}, author={Gaddy, David and Dan, Klein}, year={2022}, url={https://www2.eecs.berkeley.edu/Pubs/TechRpts/2022/EECS-2022-68.pdf}}

@misc{hifigan,
      title={HiFi-GAN: Generative Adversarial Networks for Efficient and High Fidelity Speech Synthesis}, 
      author={Jungil Kong and Jaehyeon Kim and Jaekyoung Bae},
      year={2020},
      eprint={2010.05646},
      archivePrefix={arXiv},
      primaryClass={cs.SD},
      url={https://arxiv.org/abs/2010.05646}, 
}

@article{brennan2019hierarchical,
  title={Hierarchical structure guides rapid linguistic predictions during naturalistic listening},
  author={Brennan, Jonathan R. and Hale, John T.},
  journal={PLoS ONE},
  volume={14},
  number={1},
  pages={e0207741},
  year={2019},
  publisher={Public Library of Science},
  doi={10.1371/journal.pone.0207741}
}

@article{saldanha2022data, title={Data augmentation using Variational Autoencoders for improvement of respiratory disease classification}, author={Saldanha, Jane and Chakraborty, Shaunak and Patil, Shruti and Kotecha, Ketan and Kumar, Satish and Nayyar, Anand}, journal={PLoS ONE}, volume={17}, number={8}, pages={e0266467}, year={2022}, publisher={Public Library of Science}, doi={10.1371/journal.pone.0266467} }

@INPROCEEDINGS{8282225,
  author={Nishizaki, Hiromitsu},
  booktitle={2017 Asia-Pacific Signal and Information Processing Association Annual Summit and Conference (APSIPA ASC)}, 
  title={Data augmentation and feature extraction using variational autoencoder for acoustic modeling}, 
  year={2017},
  volume={},
  number={},
  pages={1222-1227},
  doi={10.1109/APSIPA.2017.8282225}}

\vfill

\end{document}